\documentclass[%
reprint,
superscriptaddress,
showpacs,preprintnumbers,
 amsmath,amssymb,
 aps,
longbibliography,
pra,
floatfix,
]{revtex4-1}
\usepackage{float}
\usepackage{xspace}
\usepackage{graphicx}% Include figure files
\graphicspath{{./}}
\usepackage{dcolumn}% Align table columns on decimal point
\usepackage{bm}% bold math
\usepackage{color}
\usepackage{soul}
\usepackage{multirow}
\usepackage{hyperref}
 \usepackage{amssymb}
 \usepackage{nicefrac}
\hypersetup{
    unicode=false,          % non-Latin characters in Acrobat's bookmarks
    pdftoolbar=true,        % show Acrobat's toolbar?
    pdfmenubar=true,        % show Acrobat's menu?
    pdffitwindow=false,     % window fit to page when opened
    pdfstartview={FitH},    % fits the width of the page to the window
    pdftitle={Phonon anomalies in FeS},    % title
    pdfnewwindow=true,      % links in new window
    colorlinks=true,       % false: boxed links; true: colored links
    linkcolor=blue,          % color of internal links (change box color with linkbordercolor)
    citecolor=blue,        % color of links to bibliography
    filecolor=blue,      % color of file links
    urlcolor=blue,       % color of external links
    plainpages=false,        % hat Einfluss auf die Seitennummerierung
}

\usepackage{CJK}%%%%%%%%%%%%%%%
\newcommand{\Ts}{\ensuremath{T_\mathrm{s}}\xspace}

\newcommand{\Alg}{\texorpdfstring{\ensuremath{A_{1g}}\xspace}{A1g}}
\newcommand{\Ag}{\texorpdfstring{\ensuremath{A_{g}}\xspace}{Ag}}
\newcommand{\Bg}{\texorpdfstring{\ensuremath{B_{g}}\xspace}{Bg}}
\newcommand{\AZg}{\texorpdfstring{\ensuremath{A_{2g}}\xspace}{A2g}}

\newcommand{\Eg}{\texorpdfstring{\ensuremath{E_{g}}\xspace}{Eg}}
\newcommand{\Au}{\texorpdfstring{\ensuremath{A_{u}}\xspace}{Ag}}
\newcommand{\Eu}{\texorpdfstring{\ensuremath{E_{u}}\xspace}{Eu}}

\usepackage{array}
\newcolumntype{L}[1]{>{\raggedright\let\newline\\\arraybackslash\hspace{0pt}}m{#1}}

\newcommand{\wn}{\ensuremath{\rm cm^{-1}}\xspace}
\begin{document}

%%%%%%%%%%%%%%%%%%%%%%%%%%%%%%%%%%%%
%\begin{CJK*}{GBK}{}%%%%%%%%%%%%%%%%%
%%%%%%%%%%%%%%%%%%%%%%%%%%%%%%%%%%%%

\title{Lattice dynamics and phase transition in \texorpdfstring{$\mathrm{CrI_3}$}{CrI3} single crystals}
\date{\today}

\author{S. Djurdji\'{c}-Mijin}
\affiliation{Center for Solid State Physics and New Materials, Institute of Physics Belgrade, University of Belgrade, Pregrevica 118, 11080 Belgrade, Serbia}
\author{A. \v{S}olaji\'{c}}
\affiliation{Center for Solid State Physics and New Materials, Institute of Physics Belgrade, University of Belgrade, Pregrevica 118, 11080 Belgrade, Serbia}
\author{J. Pe\v{s}i\'{c}}
\affiliation{Center for Solid State Physics and New Materials, Institute of Physics Belgrade, University of Belgrade, Pregrevica 118, 11080 Belgrade, Serbia}
\author{M. \v{S}\'{c}epanovi\'{c}}
\affiliation{Center for Solid State Physics and New Materials, Institute of Physics Belgrade, University of Belgrade, Pregrevica 118, 11080 Belgrade, Serbia}
\author{Y.~Liu} %(ÁõÓý)}
\affiliation{Condensed Matter Physics and Materials Science Department, Brookhaven National Laboratory, Upton, NY 11973-5000, USA}
\author{A. Baum}
\affiliation{Walther Meissner Institut, Bayerische Akademie der Wissenschaften,
85748 Garching, Germany}
\affiliation{Fakult\"at f\"ur Physik E23, Technische Universit\"at M\"unchen, 85748 Garching, Germany}
\author{C.~Petrovic}
\affiliation{Condensed Matter Physics and Materials Science Department, Brookhaven National Laboratory, Upton, NY 11973-5000, USA}
\author{N.~Lazarevi\'{c}}
\affiliation{Center for Solid State Physics and New Materials, Institute of Physics Belgrade, University of Belgrade, Pregrevica 118, 11080 Belgrade, Serbia}
\author{Z. V.~Popovi\'{c}}
\affiliation{Center for Solid State Physics and New Materials, Institute of Physics Belgrade, University of Belgrade, Pregrevica 118, 11080 Belgrade, Serbia}
\affiliation{Serbian Academy of Sciences and Arts, Knez Mihailova 35, 11000 Belgrade, Serbia}

%%%%%%%%%%%%%%%%%%%%%%%%%%%%%%%%%%%%%%%%%%%%%%%%%%%%%%%%%%%%%%%%%%%%%%%%%%%%
\begin{abstract}
The vibrational properties of $\mathrm{CrI_3}$ single crystals were investigated using Raman spectroscopy and were analyzed with respect to the changes of the crystal structure. All but one mode are observed for both the low-temperature $R\bar{3}$ and the high-temperature C2/$m$ phase. For all observed modes the energies and symmetries are in good agreement with DFT calculations. The symmetry of a single-layer was identified as $p\bar{3}1/m$.
In contrast to previous studies we observe the transition from the $R\bar{3}$ to the $\mathrm{C2}/m$ phase at 180\,K and find no evidence for coexistence of both phases over a wide temperature range.
\end{abstract}
%%%%%%%%%%%%%%%%%%%%%%%%%%%%%%%%%%%%%%%%%%%%%%%%%%%%%%%%%%%%%%%%%%%%%%%%%%%%%%%%%%%%%%%%%%%%%%%
\maketitle

%%%%%%%%%%%%%%%%%%%%%%%%%
%\end{CJK*}%%%%%%%%%%%%%%
%%%%%%%%%%%%%%%%%%%%%%%%%

%%%%%%%%%%%%%%%%%%%%%%%%%%%%%%%%%%%%%%%%%%%%%%%%%%%%%%%%%%%%%%%%%%%%%%%%%%%%%%%%%%%%%%%%%%%%

\section{Introduction}

Two-dimensional layered materials have gained attention due to their unique properties, the potential for a wide spectrum of applications and the opportunity for the development of functional van der Waals heterostructures. $\mathrm{CrI_3}$ is a member of the chromium-trihalide family which are ferromagnetic semiconductors  \cite{B.Huang2017_N546_270-273}.
Recently, they have received significant attention as candidates for the study of magnetic monolayers. The experimental realization of  $\mathrm{CrI_3}$ ferromagnetic monolayers \cite{B.Huang2017_N546_270-273} motivated further efforts towards their understanding.
$\mathrm{CrI_3}$ features electric field controlled magnetism \cite{Jiang2018_Nn_1} as well as a strong magnetic anisotropy \cite{McGuire2015_CoM27_612-620, Lado2017_2D4_3}. With the main absorption peaks lying in the visible part of the spectrum, it is a great candidate for low-dimensional semiconductor spintronics \cite{W.-B.Zhang2015_JMCC3_12457-12468}. In its ground state, $\mathrm{CrI_3}$ is a ferromagnetic semiconductor with a Curie temperature of 61\,K \cite{Jr.1965_CM36_1259, B.Huang2017_N546_270-273} and a band-gap of 1.2\,eV \cite{Jr.1965_CM36_1259}.
It was demonstrated that the magnetic properties of $\mathrm{CrI_3}$ mono- and bilayers can be controlled by electrostatic doping \cite{Jiang2018_Nn_1}. Upon cooling, $\mathrm{CrI_3}$ undergoes a phase transition around 220\,K from the high-temperature monoclinic ($\mathrm{C2}/m$) to the low-temperature rhombohedral ($R\bar{3}$) phase \cite{McGuire2015_CoM27_612-620,Larson2018_ArXive}. Although the structural phase transition is reported to be of the first-order, it was suggested that the phases may coexist over a wide temperature range \cite{McGuire2015_CoM27_612-620}. Raman spectroscopy  can be of use here due to its capability to simultaneously probe both phases in a phase-separated system \cite{Lazarevic2012_PRB86_054503,Ryu2015_PRB91_184503,Ryu2015_PRB92_174522}.

A recent theoretical study predicted the energies of all Raman active modes in the low-temperature and high-temperature structure of $\mathrm{CrI_3}$ suggesting a near degeneracy between the \Ag and \Bg modes in the monoclinic ($\mathrm{C2}/m$) structure. Their energies match the energies of \Eg modes in the rhombohedral ($R\bar{3}$) structure \cite{Larson2018_ArXive}.

In this article we present an experimental and theoretical Raman scattering study of $\mathrm{CrI_3}$ lattice dynamics. In both phases all but one of the respective modes predicted by symmetry were observed. The energies for all modes are in good agreement with the theoretical predictions for the assumed crystal symmetry. Our data suggest that the first-order transition occurs at $\Ts \approx 180$\,K without evidence for phase coexistence over a wide temperature range.

%%%%%%%%%%%%%%%%%%%%%%%%%%%%%%%%%%%%%%%%%%
\section{Experiment and numerical method}

The preparation of the single crystal $\mathrm{CrI_3}$ sample used in this study is described elsewhere \cite{Liu2018_PRB97_014420}. The Raman scattering experiment was performed using a Tri Vista 557 spectrometer in backscattering micro-Raman configuration with a 1800/1800/2400\,groves/mm diffraction grating combination. The 532\,nm line of a Coherent Verdi G solid state laser was used as an excitation source. The direction of the incident light coincides with the crystallographic $c$ axis. The sample was oriented so that its principal axis of the $R\bar{3}$ phase coincides with the $x$ axis of the laboratory system.  A KONTI CryoVac continuous Helium flow cryostat with a 0.5\,mm thick window was used for measurements at all temperatures under high vacuum ($10^{-6}$\,mbar). The sample was cleaved in air before being placed into the cryostat. The obtained Raman spectra were corrected by the Bose factor and analysed quantitatively by fitting Voigt profiles to the data whereby the Gaussian width $\Gamma_{\mathrm{Gauss}} = 1\,\wn$ reflects the resolution of the spectrometer.

The spin polarized density functional theory (DFT) calculations have been performed in the Quantum Espresso (QE) software package \cite{QE-2009}, using the Perdew-Burke-Ernzehof (PBE) exchange-correlation functional \cite{Perdew1996_PRL77_3865} and PAW pseudopotentials \cite{Bloechl1994_PRB50_17953--17979,Kresse1999_PRB59_1758--1775}. The energy cutoffs for the wavefunctions and the charge density were set to be 85\,Ry and 425\,Ry, respectively, after convergence tests. For $k$-point sampling, the Monkhorst-Pack scheme was used with a $8\times8\times8$ grid centered around the $\Gamma$ point. Optimization of a atomic positions in the unit cell was performed until the interatomic forces were smaller than $10^{-6}\,\mathrm{Ry/\mathring{A}}$. To treat the van der Waals (vdW) interactions a Grimme-D2 correction \cite{Stefan_JoCC27_1787-1799} is used in order to include long-ranged forces between the layers, which are not properly captured within LDA or GGA functionals. This way, the parameters are obtained more accurately, especially the interlayer distances. Phonon frequencies were calculated at the $\Gamma$ point using the linear response method implemented in QE. The phonon energies are compiled in Table~\ref{ref:Table3} together with experimental values. Eigenvectors of Raman active modes for both low- and high-temperature phases are depicted in Figure~\ref{ref:FigureA1} of the Appendix.

\section{Results and Discussion}

%%%%%%%%%%%%%%%%%%%%%%%%%%%%%%%%%%%%%%%%%%%%%%%%%%%%%%%%%%%%%%%%%%%%%%
\begin{figure}
 \centering
  \includegraphics[width=85mm]{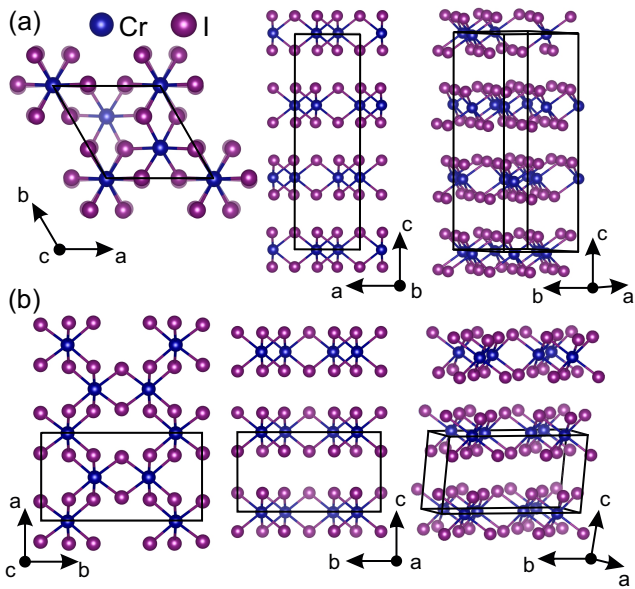}
  \caption{(Colour online) Schematic representation of (a) the low-temperature $R\bar{3}$ and (b) the high-temperature $\mathrm{C2}/m$ crystal structure of $\mathrm{CrI_3}$. Black lines represent unit cells.}
 \label{ref:Figure1}
\end{figure}
%%%%%%%%%%%%%%%%%%%%%%%%%%%%%%%%%%%%%%%%%%%%%%%%%%%%%%%%%%%%%%%%%%%%%%

%%%%%%%%%%%%%%%%%%%%%%%%%%%%%%%%%%%%%%%%%%%%%%%%%%%%%%%%%%%%%%%%%%%%%%
\begin{table}[t]
\caption{Calculated and experimental \cite{Liu2018_PRB97_014420} parameters of the crystallographic unit cell for the low-temperature $R\bar{3}$ and high-temperature  $\mathrm{C2}/m$ phase of $\mathrm{CrI_3}$.}
\label{ref:Table1}
\begin{ruledtabular}
\centering
\resizebox{\linewidth}{!}{%
\begin{tabular}{ccccc}
\multirow{2}{*}{$T(\mathrm{K})$} &\multicolumn{2}{c} {Space group  $R\bar{3}$} &\multicolumn{2}{c}{Space group $\mathrm{C2}/m$ }\\ \cline{2-3} \cline{4-5} \\[-1mm]
 &  Calc. & Exp. \cite{Liu2018_PRB97_014420}  & Calc. & Exp. \cite{Liu2018_PRB97_014420} \\[1mm] \hline \\[-0.5em]
$a~(\mathrm{\mathring{A}})$ &6.87& 6.85  & 6.866 & 6.6866\\[1mm]
$b~(\mathrm{\mathring{A}})$ & 6.87 & 6.85 & 11.886 & 11.856 \\[1mm]
$c~(\mathrm{\mathring{A}})$ & 19.81 & 19.85 & 6.984 & 6.966  \\[1mm]
$\mathrm{\alpha~(deg)}$ & 90 & 90 & 90 & 90 \\[1mm]
$\mathrm{\beta~(deg)}$  & 90 & 90&108.51&108.68\\[1mm]
$\mathrm{\gamma~(deg)}$ & 120 & 120&90 &90 \\[1mm]

\end{tabular}}
\end{ruledtabular}
\end{table}
%%%%%%%%%%%%%%%%%%%%%%%%%%%%%%%%%%%%%%%%%%%%%%%%%%%%%%%%%%%%%%%%%%%%%%

%%%%%%%%%%%%%%%%%%%%%%%%%%%%%%%%%%%%%%%%%%%%%%%%%%%%%%%%%%%%%%%%%%%%%%
\begin{figure}
 \centering
  \includegraphics[width=85mm]{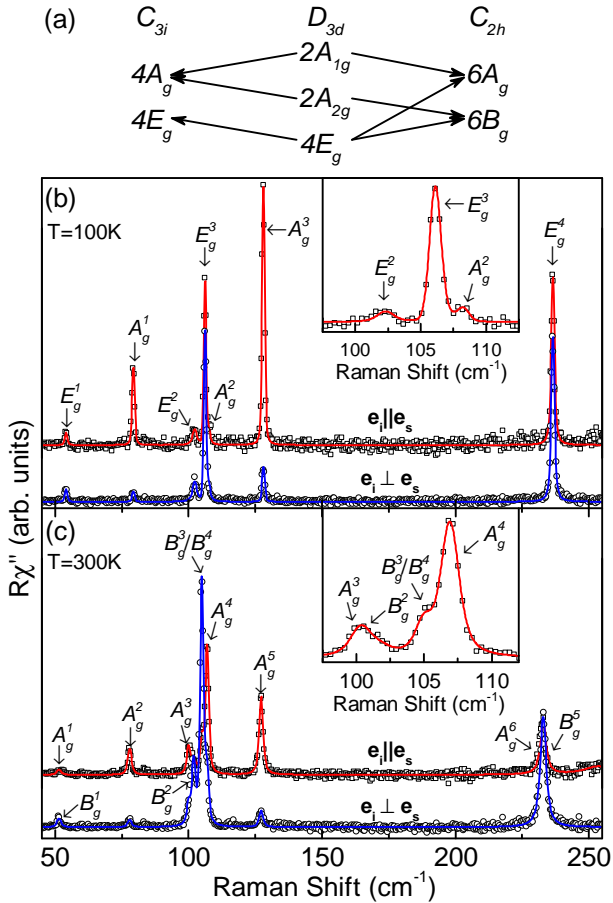}
  \caption{(Colour online) (a) Compatibility relations for the $\mathrm{CrI_3}$ layer and the crystal symmetries. Raman spectra of (b) the low-temperature $R\bar{3}$ and (c) the high-temperature $\mathrm{C2}/m$ crystal structure measured in parallel (open squares) and crossed (open circles) polarization configurations at 100\,K and 300\,K, respectively. Red and blue solid lines represent fits of Voigt profiles to the experimental data.}
 \label{ref:Figure2}
\end{figure}

%%%%%%%%%%%%%%%%%%%%%%%%%%%%%%%%%%%%%%%%%%%%%%%%%%%%%%%%%%%%%%%%%%%%%

%%%%%%%%%%%%%%%%%%%%%%%%%%%%%%%%%%%%%%%%%%%%%%%%%%%%%%%%%%%%%%%%%%%%%%
\begin{table*}[t]
\caption{Wyckoff positions of the two types of atoms and their contributions to the $\Gamma$-point phonons for the $R\bar{3}$ and $\mathrm{C2}/m$ as well as the $p\bar{3}1/m$ diperiodic space group. The second row shows the Raman tensors for the corresponding space groups.}
\label{ref:Table2}
\begin{ruledtabular}
\centering
%\resizebox{\linewidth}{!}{%
\begin{tabular}{ccccccccc }
\multicolumn{3}{c} {Space group $R\bar{3}$} &\multicolumn{3}{c}{Diperiodic space group $p\bar{3}1/m$}&\multicolumn{3}{c}{Space group: $\mathrm{C2}/m$}\\ \cline{1-3} \cline{4-6} \cline{7-9} \\[-2mm]

Atoms &  \multicolumn{2}{c} {Irreducible representations} & Atoms & \multicolumn{2}{c} {Irreducible representations} & Atoms &  \multicolumn{2}{c} {Irreducible representations} \\ \cline{1-3} \cline{4-6} \cline{7-9} \\[-2mm]

Cr ($6c$) &  \multicolumn{2}{c} {\Ag + \Au + \Eg+ \Eu} & Cr ($2c$) &  \multicolumn{2}{c} {\AZg +$A_{2u}$ + \Eg+ \Eu }& Cr ($4g$) & \multicolumn{2}{c} {$A_{g} + A_{u} + 2B_{g}+ 2B_{u}$}\\[1mm]

\multirow{2}{*}{I (18f)} &  \multicolumn{2}{c} {\multirow{2}{*}{3\Ag + 3\Au + 3\Eg+ 3\Eu}} & \multirow{2}{*}{I ($6k$)} & \multicolumn{2}{c} { $2A_{1g}+A_{1u}+A_{2g}+$} & I ($4i$) & \multicolumn{2}{c} {$2A_{g} + 2A_{u} + B_{g}+ B_{u}$} \\

\multirow{2}{*}{}&\multicolumn{2}{c} {\multirow{2}{*}{}}&\multirow{2}{*}{}&\multicolumn{2}{c} {$+2A_{2u}+3E_{g}+3E_{u}$}& I ($8j$) & \multicolumn{2}{c} {3$A_{g} + 3A_{u} + 3B_{g}+3B_{u}$}\\

\cline{1-3} \cline{4-6} \cline{7-9} \\[-2mm]

\multicolumn{3}{c}{$
\Ag = \begin{pmatrix}
a& & \\
 &a& \\
 & &b\\\end{pmatrix}
$}
&
\multicolumn{3}{c}{
$\Alg = \begin{pmatrix}
a& & \\
 &a& \\
 & &b\\
\end{pmatrix}
$}&
\multicolumn{3}{c}{
$\Ag = \begin{pmatrix}
a& &d\\
 &c& \\
d& &b\\
\end{pmatrix}
$}
\\ [5mm]

\multicolumn{3}{c}{$
{}^1\Eg = \begin{pmatrix}
c&d&e\\
d&-c&f\\
e&f& \\ \end{pmatrix}
\;
{}^2\Eg = \begin{pmatrix}
d&-c&-f\\
-c&-d&e\\
-f&e& \\
\end{pmatrix}
$}&
\multicolumn{3}{c}{$
{}^1\Eg = \begin{pmatrix}
c&  & \\
 &-c&d\\
 & d& \\ \end{pmatrix}
\;
{}^2\Eg = \begin{pmatrix}
  &-c&-d\\
-c&  &  \\
-d& e&  \\
\end{pmatrix}
$}
&
\multicolumn{3}{c}{$
\Bg = \begin{pmatrix}
 &e& \\
e& &f\\
 &f& \\ \end{pmatrix}$
}
\\ [5mm]

\end{tabular}
\end{ruledtabular}
\end{table*}
%%%%%%%%%%%%%%%%%%%%%%%%%%%%%%%%%%%%%%%%%%%%%%%%%%%%%%%%%%%%%%%%%%%%%

\begin{table*}[t]
\caption{Phonon symmetries and phonon energies for the low-temperature $R\bar{3}$ and high-temperature $\mathrm{C2}/m$ phase of $\mathrm{CrI_3}$. The experimental values were determined at 100\,K and 300\,K, respectively. All calculations were performed at zero temperature. Arrows indicate the correspondence of the phonon modes across the phase transition.}
\label{ref:Table3}
\begin{ruledtabular}
\centering
%\resizebox{\linewidth}{!}{%
\begin{tabular}{*{9}{c}}
\multicolumn{4}{c} {Space group  $R\bar{3}$} & &\multicolumn{4}{c}{Space group $\mathrm{C2}/m$ }\\ \cline{1-4} \cline{6-9} \\[-2mm]

Symm. & Exp. (\wn) & Calc. (\wn) & Calc. (\wn) \cite{Larson2018_ArXive} & & Symm. & Exp. (\wn) & Calc. (\wn) & Calc. \cite{Larson2018_ArXive} (\wn) \\ [1mm] \cline{1-4} \cline{6-9}  \\[-0.5em]

\multirow{ 2}{*}{$\Eg^{1}$}  & \multirow{ 2}{*}{54.1} & \multirow{ 2}{*}{59.7} & \multirow{ 2}{*}{53} & \multirow{ 2}{*}{{\begin{tabular}{c}\rotatebox[origin=l]{-10}{$\xrightarrow{\hspace*{5mm}}$} \\[-5mm] \rotatebox[origin=l]{10}{$\xrightarrow{\hspace*{5mm}}$} \end{tabular}}} & $\Bg^{1}$ & 52.0 & 57.0 & 52\\[1mm] %\hline \\[-0.5em]
    &  & & &  &  $\Ag^{1}$ & 53.6  & 59.8 & 51 \\[1mm]

$\Ag^{1}$ & 73.33 & 89.6 & 79 & $\xrightarrow{\makebox[5mm]{}}$ & $\Ag^{2}$ & 78.6 & 88.4 & 79 \\[1mm]

\multirow{ 2}{*}{$\Eg^{2}$} &  \multirow{ 2}{*}{102.3} & \multirow{ 2}{*}{99.8} & \multirow{ 2}{*}{98}  & \multirow{ 2}{*}{{\begin{tabular}{c}\rotatebox[origin=l]{-10}{$\xrightarrow{\hspace*{5mm}}$} \\[-5mm] \rotatebox[origin=l]{10}{$\xrightarrow{\hspace*{5mm}}$} \end{tabular}}} & $\Ag^{3}$ & 101.8 & 101.9 & 99 \\[1mm]
  & &  &  & &  $\Bg^{2}$ & 102.4 & 101.8 & 99 \\[1mm]

\multirow{ 2}{*}{$\Eg^{3}$} & \multirow{ 2}{*}{106.2} & \multirow{ 2}{*}{112.2} & \multirow{ 2}{*}{102}  & \multirow{ 2}{*}{{\begin{tabular}{c}\rotatebox[origin=l]{-10}{$\xrightarrow{\hspace*{5mm}}$} \\[-5mm] \rotatebox[origin=l]{10}{$\xrightarrow{\hspace*{5mm}}$} \end{tabular}}} & $\Bg^{3}$ & 106.4* & 108.9 & 101 \\[1mm]
  & &  &   & & $\Ag^{4}$ & 108.3 & 109.3 & 102 \\[1mm]

$\Ag^{2}$ & 108.3 & 98.8 & 88 & $\xrightarrow{\makebox[5mm]{}}$ & $\Bg^{4}$ & 106.4* & 97.8 & 86 \\[1mm]

$\Ag^{3}$ & 128.1 & 131.1 & 125 & $\xrightarrow{\makebox[5mm]{}}$ & $\Ag^{5}$ & 128.2 & 131.7 & 125 \\[1mm]

$\Ag^{4}$ & - & 195.2 & 195 & $\xrightarrow{\makebox[5mm]{}}$ & $\Bg^{5}$ & - & 198.8 & 195 \\[1mm]

\multirow{ 2}{*}{$\Eg^{4}$} & \multirow{ 2}{*}{236.6}  & \multirow{ 2}{*}{234.4} & \multirow{ 2}{*}{225}  & \multirow{ 2}{*}{{\begin{tabular}{c}\rotatebox[origin=l]{-10}{$\xrightarrow{\hspace*{5mm}}$} \\[-5mm] \rotatebox[origin=l]{10}{$\xrightarrow{\hspace*{5mm}}$} \end{tabular}}} & $\Ag^{6}$ & 234.6 & 220.1 & 224 \\[1mm]
 &   &  &   &  & $\Bg^{6}$ & 235.5 & 221.1 & 225 \\[0mm]
\end{tabular}
\end{ruledtabular}
* observed as two peak structure \hspace{130mm}
\end{table*}
%%%%%%%%%%%%%%%%%%%%%%%%%%%%%%%%%%%%%%%%%%%%%%%%%%%%%%%%%%%%%%%%%%%%%

$\mathrm{CrI_3}$ adopts a rhombohedral $R\bar{3}$ ($C_{3i}^2$) crystal structure at low temperatures and a monoclinic $\mathrm{C2}/m$ ($C_{2h}^3$) crystal structure at room temperature \cite{McGuire2015_CoM27_612-620}, as shown in Figure~\ref{ref:Figure1}.  The main difference between the high- and low-temperature crystallographic space groups arises from different stacking sequences,  with  $\mathrm{CrI_3}$ layers being almost identical. In the rhombohedral structure the Cr atoms in one layer are placed above the center of a hole in the Cr honeycomb net of the two adjacent layers. When crossing the structural phase transition at \Ts to the monoclinic structure the layers are displaced along the $a$ direction so that every fourth layer is at the same place as the first one. The interatomic distances, mainly the interlayer distance, and the vdW gap, are slightly changed by the structural transition. The crystallographic parameters for both phases are presented in Table~\ref{ref:Table1}. The numerically obtained values are in good agreement with reported X-ray diffraction data \cite{Liu2018_PRB97_014420}.

The vibrational properties of layered materials are typically dominated by the properties of single layers composing the crystal. The symmetry of a single layer can be described by one of the 80 diperiodic space groups (DG) obtained by lifting translational invariance in the direction perpendicular to the layer \cite{WoodElizabeth_BSTJ43_541-559}. In the case of $\mathrm{CrI_3}$, the symmetry analysis revealed that the single layer structure is fully captured by the $p\bar{3}1/m$ ($D_{3d}^1$) diperiodic space group DG71, rather than by $R\bar{3}2/m$ as proposed in Ref.~\cite{Larson2018_ArXive}.

%%%%%%%%%%%%%%

According to the factor group analysis (FGA) for the single $\mathrm{CrI_3}$ layer, six modes ($2\Alg+4\Eg$) are expected to be observed in the Raman scattering experiment (see Table~\ref{ref:Table2}). By stacking the layers the symmetry is reduced and, depending on the stacking sequence, FGA yields a total of eight Raman active modes ($4\Ag+4\Eg$) for the $R\bar{3}$ and twelve Raman active modes ($6\Ag+6\Bg$) for the $\mathrm{C2}/m$ crystal symmetry. The correlation between layer and crystal symmetries for both cases is shown in Figure~\ref{ref:Figure2}\,(a) \cite{G.Fateley1971_25_155-173,Lazarevic2011_PRB83_024302}.

Fig.~\ref{ref:Figure2}(b) shows the $\mathrm{CrI_3}$ single crystal Raman spectra measured at 100\,K in two scattering channels. According to selection rules for the rhombohedral crystal structure (Table~\ref{ref:Table2}) the $\Ag$ modes can be observed only in the parallel polarization configuration, whereas the $\Eg$ modes appear in both parallel and crossed polarization configurations. Based on the selection rules, peaks at about 78\,\wn, 108\,\wn and 128\,\wn were identified as $\Ag$ symmetry modes, whereas  peaks at about  54\,\wn, 102\,\wn, 106\,\wn and 235\,\wn are assigned as \Eg symmetry. The weak observation of the most pronounced $\Ag$ modes in crossed polarisations [Fig.~\ref{ref:Figure2}\,(b)] is attributed to the leakage due to a slight sample misalignment and/or the presence of defects in the crystal.
The energies of all observed modes are compiled in Table~\ref{ref:Table3} together with the energies predicted by our calculations and by Ref.~\cite{Larson2018_ArXive} and are found to be in good agreement for the \Eg modes. The discrepancy is slightly larger for the low energy \Ag modes. Our calculations in general agree with those from Ref.~\cite{Larson2018_ArXive}.
The $\Ag^4$ mode of the rhombohedral phase, predicted by calculation to appears at about 195 \wn, was not observed in the experiment, most likely due to its low intensity.

When the symmetry is lowered in the high-temperature monoclinic $\mathrm{C}2/m$ phase [Fig.~\ref{ref:Figure2}(c)] the \Eg modes split into an \Ag and a \Bg mode each, whereas the rhombohedral $\Ag^2$ and $\Ag^4$ modes are predicted to switch to the monoclinic \Bg symmetry. The correspondence of the phonon modes across the phase transition is indicated by the arrows in Table~\ref{ref:Table3}. The selection rules for C2/$m$ (see Table~\ref{ref:Table2}) predict that \Ag and \Bg modes can be observed in both parallel and crossed polarization configurations. Additionally, the sample forms three types of domains which are rotated with respect to each other.
We therefore identify the phonons in the C2/$m$ phase in relation to the calculations and find again good agreement of the energies. The $\Bg^3$ and $\Bg^4$ modes overlap and therefore cannot be resolved separately. As can be seen from the temperature dependence shown below [Fig.~\ref{ref:Figure4}(b)] the peak at 106\,\wn broadens and gains spectral weight in the monoclinic phase in line with the expectance that two modes overlap. The missing rhombohedral $\Ag^4$ mode corresponds to the monoclinic $\Bg^5$ mode, which is likewise absent in the spectra.

%%%%%%%%%%%%%%&&&&&&&&&&&&&&&&&&&&&&&&&&&&&&&&&&&&&&&&&&&&&&&&&&&&&&&&
\begin{figure}
 \centering
  \includegraphics[width=85mm]{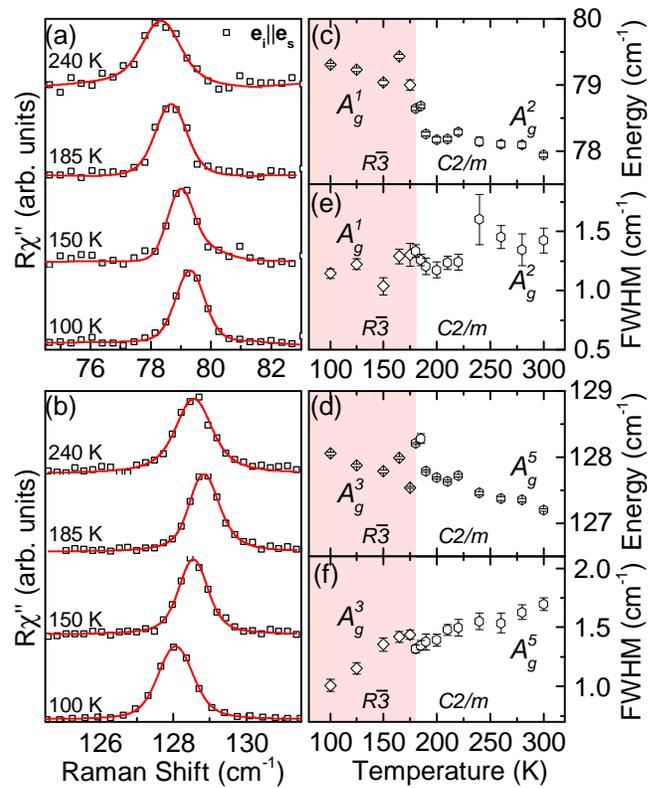}
  \caption{(Colour online) Temperature dependence of the $\Ag^1$ and $\Ag^3$ phonon modes of the rhombohedral structure and the corresponding $\Ag^2$ and $\Ag^5$ modes of the monoclinic structure, respectively. (a,b) Raman spectra at temperatures as indicated. The spectra are shifted for clarity. Solid red lines represent Voigt profiles fitted to the data. (c,d) and (e,f) Temperature dependence of the phonon energies and line widths, respectively. Both modes show an abrupt change in energy at the phase transition at 180\,K. }
 \label{ref:Figure3}
\end{figure}
%%%%%%%%%%%%%%%%%%%%%%%%%%%%%%%%%%%%%%%%%%%%%%%%%%%%%%%%%%%%%%%%%%%%%%

%%%%%%%%%%%%%%%%%%%%%%%%%%%%%%%%%%%%%%%%%%%%%%%%%%%%%%%%%%%%%%%%%%%%%%%%%%%%%%%%%%%%
\begin{figure}
 \centering
  \includegraphics[width=85mm]{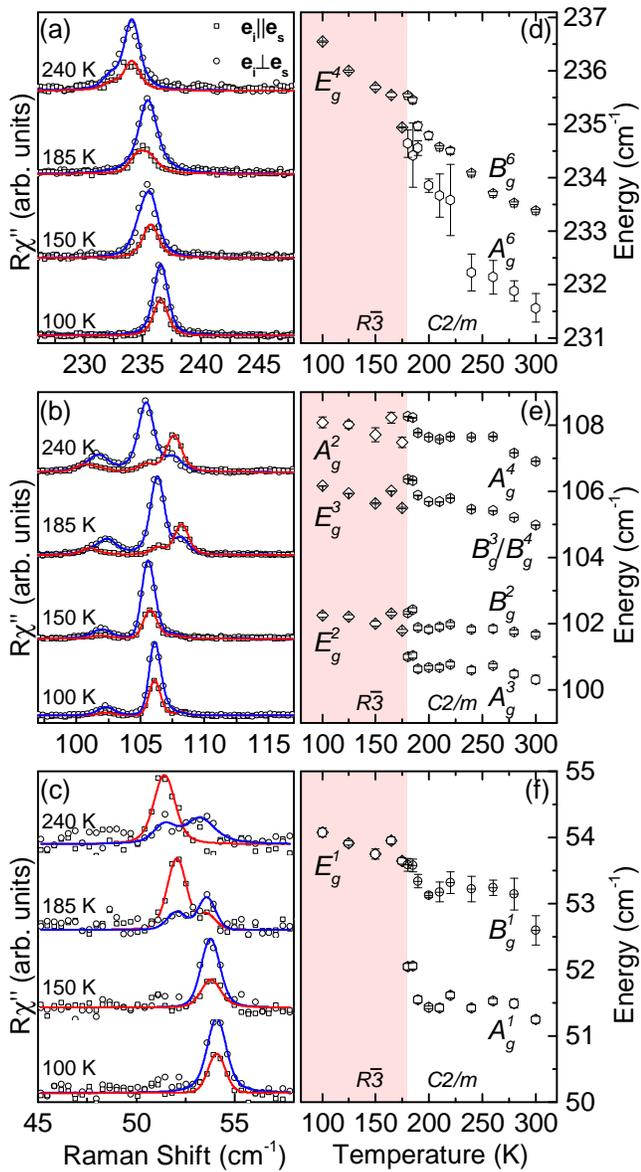}
  \caption{(Colour online) Temperature dependence of the rhombohedral $\Ag^4$ and \Eg modes. (a-c) Raman spectra in parallel (open squares) and crossed (open circles) light polarizations at temperatures as indicated. The spectra are shifted for clarity. Blue and red solid lines are fits of Voigt profiles to the data. Two spectra were analysed simultaneously in two scattering channels with the integrated intensity as the only independent parameter. (d-f) Phonon energies obtained from the Voigt profiles. Each \Eg mode splits into an \Ag and a \Bg mode above 180\,K. }
 \label{ref:Figure4}
\end{figure}
%%%%%%%%%%%%%%%%%%%%%%%%%%%%%%%%%%%%%%%%%%%%%%%%%%%%%%%%%%%%%%%%%%%%%

The temperature dependence of the observed phonons is shown in Figs.~\ref{ref:Figure3} and \ref{ref:Figure4}. In the low-temperature rhombohedral phase all four \Eg modes as well as $\Ag^1$ and $\Ag^2$ soften upon warming, whereas $\Ag^3$ hardens up to $T \approx 180\,\mathrm{K}$ before softening again.
Crossing the first order phase transition from $R\bar{3}$ to $\mathrm{C2}/m$ crystal symmetry is reflected in the spectra as a symmetry change and/or renormalization for the non-degenerate modes and lifting of the degeneracy of the \Eg modes as shown in Table~\ref{ref:Table2}. In our samples, this transition is observed at $\Ts \approx 180\,\mathrm{K}$. The splitting of the \Eg phonons into \Ag and \Bg modes at the phase transition is sharp [Fig.~\ref{ref:Figure4}]. The rhombohedral $\Ag^1$ and $\Ag^3$ phonons show a jump in energy and a small discontinuity in the line width at \Ts [Fig.~\ref{ref:Figure3}].
Our spectra were taken during warming in multiple runs after cooling to 100\,K each time. We found that the temperature dependence for the phonon modes obtained this way was smooth in each phase.
McGuire \textit{et al.} \cite{McGuire2015_CoM27_612-620,McGuire_PRM1_014001} reported \Ts in the range of 220\,K, a coexistence of both phases and a large thermal hysteresis. However, they also noted that the first and second warming cycle showed identical behaviour and only found a shift of the transition temperature to higher values of the cooling cycles.
We therefore consider the difference between the reported transition around 220\,K and our $\Ts \approx 180\,\mathrm{K}$ significant. To some extent this difference may be attributed to local heating by laser.
More importantly, we find no signs of phase coexistence in the observed temperature range. The spectra for the low-temperature and high-temperature phases are distinctly different [Fig.~\ref{ref:Figure2}] and the \Eg modes exhibit a clearly resolved splitting which occurs abruptly at \Ts.
We performed measurements in the small temperature steps (see Figs.~\ref{ref:Figure3} and \ref{ref:Figure4}). This limits the maximum temperature interval where the phase coexistence could occur to approximately 5\,K, much less than the roughly 30 to 80\,K reported earlier \cite{McGuire2015_CoM27_612-620,McGuire_PRM1_014001}.

\section{Conclusion}
We studied the lattice dynamics in the single crystalline $\mathrm{CrI_3}$ using Raman spectroscopy supported by numerical calculations. For both the low-temperature $R\bar{3}$ and the high-temperature C2/$m$ phases, all except one of the predicted phonon modes were identified and the calculated and experimental phonon energies were found to be in good agreement. We determined that the symmetry of the single $\mathrm{CrI_3}$ layers is $p\bar{3}1/m$.
Abrupt changes to the spectra were found at the first-order phase transition which was located at $\Ts \approx 180\,\mathrm{K}$, lower than in previous studies. In contrast to the prior reports we found no sign of phase coexistence over temperature ranges exceeding 5\,K.

\section*{Acknowledgement}
The work was supported by the Serbian Ministry of Education, Science and Technological Development under Projects III45018 and OI171005. DFT calculations were performed using computational resources at Johannes Kepler University, Linz, Austria. Work at Brookhaven is supported by the U.S. DOE under Contract No. DESC0012704.


%\end{document}

%%%%%%%%%%%%%%%%%%%%%%%%%%%%%%%%%%%%%%%%%%%%%%%%

%\clearpage
\begin{appendix}
\label{sec:appendix}

\setcounter{figure}{0}
\renewcommand\thefigure{A\arabic{figure}}

\setcounter{table}{0}
\renewcommand\thetable{A\Roman{table}}

\section{Eigenvectors}
\label{Asec:eigenvectors}

\begin{figure*}[t]
 \centering
  \includegraphics[width=170mm]{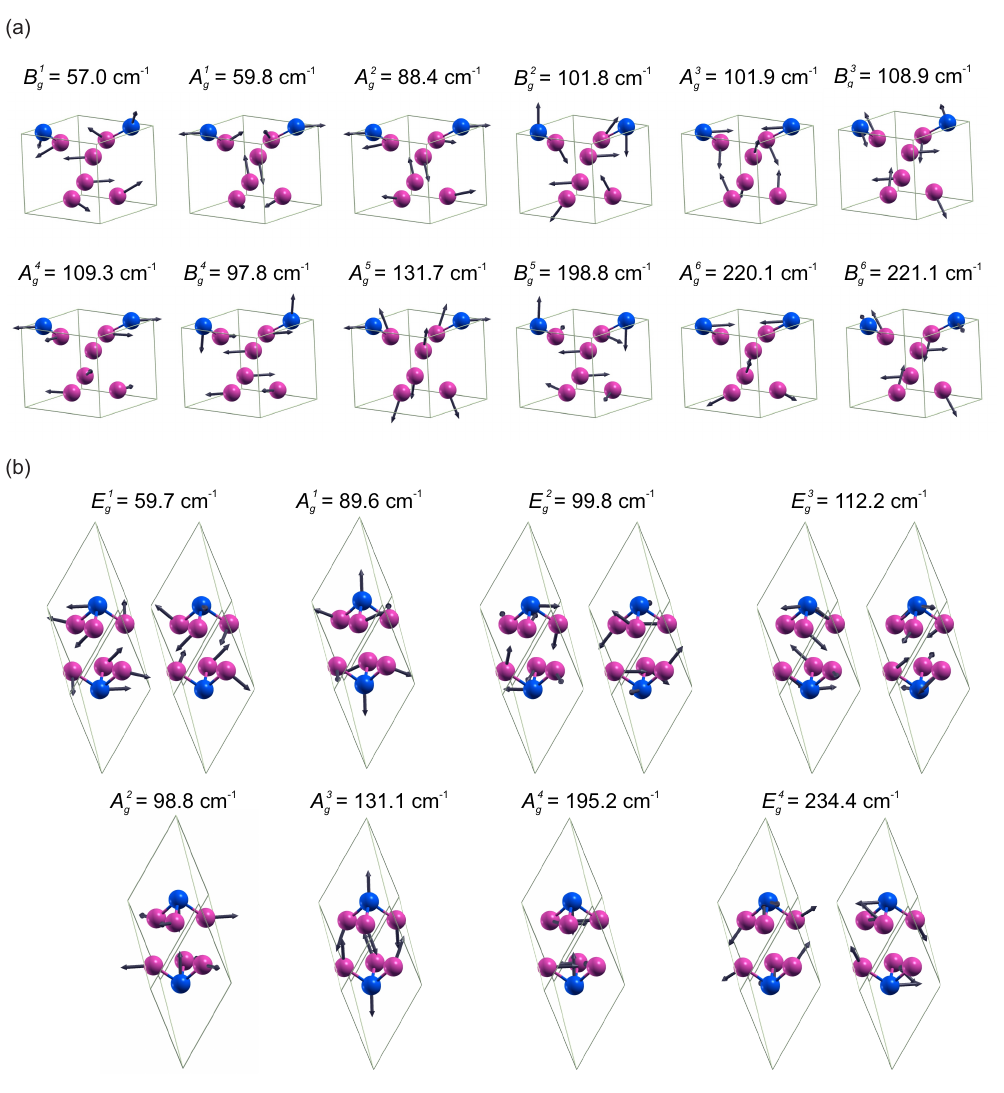}
  \caption{ Raman-active phonons in $\mathrm{CrI_3}$ for (a) the monoclinic phase hosting \Ag and \Bg modes and for (b) the rhombohedral phase hosting \Ag and \Eg modes. Blue and violet spheres denote Cr and I atoms, respectively. Solid lines represent primitive unit cells. Arrow lengths are proportional to the square root of the inter-atomic forces. The given energies are calculated for zero temperature.}
 \label{ref:FigureA1}
\end{figure*}

In addition to the phonon energies we also calculated the phonon eigenvectors which are shown in Fig.~\ref{ref:FigureA1}(a) for the high-temperature monoclinic phase and in Fig.~\ref{ref:FigureA1}(b) for the low-temperature rhombohedral phase. The energies, as given, are calculated for zero temperature. The relative displacement of the atoms is denoted by the length of the arrows.

%%%%%%%%%%%%%%%%%%%%%%%%%%%%%%%%%%%%%%%%%%%%%%%%%%%%%%%%

%%%%%%%%%%

\end{appendix}

\end{document}